\pdfoutput=1
\documentclass{sig-alternate}
\usepackage{mathptmx} 
\usepackage{kotex}
\usepackage{multirow}
\usepackage[table,xcdraw]{xcolor}
\usepackage{colortbl}
\usepackage{hhline}
\usepackage{amsmath}
\usepackage{tabto}

\usepackage{fancyhdr}
\usepackage[normalem]{ulem}
\usepackage[hyphens]{url}
\usepackage[sort,nocompress]{cite}
\usepackage[final]{microtype}
\usepackage[bookmarks=true,breaklinks=true,letterpaper=true,colorlinks,linkcolor=black,citecolor=blue,urlcolor=black]{hyperref}

\fancypagestyle{firstpage}{
  \fancyhf{}
  
  \fancyfoot[C]{\thepage}
}

\title{FMMU: A Hardware-Automated Flash Map Management Unit for Scalable Performance of NAND Flash-Based SSDs} 

\def\sharedaffiliation{%
\end{tabular}
\begin{tabular}{c}}
\begin{document}

\numberofauthors{2}
    \author{
      \alignauthor Yeong-Jae Woo
      \alignauthor Sang Lyul Min
      \sharedaffiliation
      \affaddr{\textit{Department of Computer Science and Engineering, Seoul National University}}  \\
      \affaddr{\textit{\{yjwoo, symin\}@snu.ac.kr}}
    }

\maketitle
\thispagestyle{firstpage}
\pagestyle{plain}


\begin{abstract}

NAND flash-based Solid State Drives (SSDs), which are widely used from embedded systems to enterprise servers, are enhancing performance by exploiting the parallelism of NAND flash memories. To cope with the performance improvement of SSDs, storage systems have rapidly adopted the host interface for SSDs from Serial-ATA, which is used for existing hard disk drives, to high-speed PCI express. Since NAND flash memory does not allow in-place updates, it requires special software called Flash Translation Layer (FTL), and SSDs are equipped with embedded processors to run FTL. Existing SSDs increase the clock frequency of embedded processors or increase the number of embedded processors in order to prevent FTL from acting as bottleneck of SSD performance, but these approaches are not scalable.



This paper proposes a hardware-automated Flash Map Management Unit, called \textit{FMMU}, that handles the address translation process dominating the execution time of the FTL by hardware automation. FMMU provides methods for exploiting the parallelism of flash memory by processing outstanding requests in a non-blocking manner while reducing the number of flash operations. The experimental results show that the FMMU reduces the FTL execution time in the map cache hit case and the miss case by 44\% and 37\%, respectively, compared with the existing software-based approach operating in 4-core. FMMU also prevents FTL from acting as a performance bottleneck for up to 32-channel, 8-way SSD using PCIe 3.0 x32 host interface.



\end{abstract}

\section{Introduction}

Although NAND flash-based storage devices are still more expensive per-unit-capacity than Hard Disk Drives (HDDs), they are widely used in various storage systems due to the inherent advantages of NAND flash memory such as small size, low power consumption, fast random access, noiseless, and shock resistance. In mobile environments where requirements such as small size, low power consumption, and shock resistance are critical, flash-based storage devices such as embedded Multi Media Card (eMMC) and Universal Flash Storage (UFS), are completely replacing HDDs. In datacenter and enterprise server environments, Solid State Drives (SSDs) are gradually replacing HDDs because they provide higher performance and lower total cost of ownership (TCO) of storage systems. This trend is likely to continue because the price-per-capacity of SSDs is reducing with the advert of Multi-Level Cell (MLC) and 3D NAND flash memory technology.



SSDs support a block device interface, which is the same interface of HDDs, for drop-in replacement of HDDs in existing storage systems. The block device interface assumes that the storage media allows read and write operation in sector units. Unlike disk, NAND flash memory allows read and program operation in page units, and erase operation must be performed before re-programming pages of the block. To provide the block device interface while hiding this mismatch, SSDs require a special software, called Flash Translation Layer (FTL). The FTL converts host requests of read and write requests into NAND flash memory operations of read, program, and erase operations. SSDs are equipped with embedded processors such as ARM and ARC to execute the FTL.



The performance of SSDs is continuously improving as the performance of NAND flash memories and host interfaces are improved. NAND flash operation time is divided into data transfer time for transferring data from/to page buffer of the flash memory, and cell operation time for reading, programming, and erasing data from/to the flash memory cell. The data transfer time is shortened as NAND technology evolves, and can be improved by transferring data in parallel for each channel of the multi-channel flash controller. Unfortunately the cell operation time is getting worse as NAND technology evolves, but it can be improved by using special operations such as multi-plane operation and one-shot program operation~\cite{oneshot} and by exploiting the flash memory parallelism. The performance of SSDs has already surpassed the bandwidth limitations of Serial ATA (SATA) and Serial Attached SCSI (SAS), which are conventional HDD-based interfaces. Thus, recent storage systems are adopting PCI express (PCIe) over NVM express (NVMe)~\cite{nvme} as the host interface to cope with the performance of SSDs.



As the performance of NAND flash memories and host interfaces increase, the FTL executed by embedded processors can act as an SSD performance bottleneck. The traditional way to reduce the FTL execution time is to increase the CPU clock frequency or increase the number of CPUs to run the FTL in parallel. Since power consumption increases sharply as CPU clock rate increases, SSDs employ several CPUs operating at a certain frequency. However, increasing the number of CPUs is not scalable because of the increased chip size, power consumption, and synchronization overhead among CPUs.



In this paper, we propose a hardware-automated Flash Map Management Unit, called \textit{FMMU}. FMMU handles the address translation process, which accounts for most of the FTL execution time, faster than software implemented map cache units by hardware automation. Even though the performance of NAND flash memories and host interfaces increases, FMMU prevents FTL execution time from becoming an SSD performance bottleneck. The contributions of this paper can be summarized as follows:



\begin{itemize}

\item By modeling and analyzing the performance of SSD, we identify which component of the SSD can be a future performance bottleneck.



\item We design FMMU, a hardware-automated map management unit, to reduce the FTL execution time. FMMU is designed to handle concurrent host and GC requests in a non-blocking manner, maximize NAND flash memory parallelism, and minimize the number of flash operations.



\item We evaluate the performance of SSD using FMMU based on the DiskSim with SSD extension environment to see it can scale up the SSD performance as the performance of NAND flash memory and host interface increase.



\end{itemize}

The remainder of this paper is organized as follows: Section~\ref{sec:background} introduces the background. The motivation of this paper is described in Section~\ref{sec:motivation} by analyzing the performance of SSDs. The design and implementation of the FMMU are presented in Section~\ref{sec:design}. Section~\ref{sec:evaluation} shows the performance evaluation of FMMU. Section~\ref{sec:related} discusses related studies, and the paper is concluded in Section~\ref{sec:conclusion}.



\section{Background and Motivation} \label{sec:background}

\subsection{NAND Flash Memory}

NAND flash memory is mainly used as a storage medium because it is a non-volatile memory that retains data even when power is not supplied. A single NAND flash memory chip consists of thousands of blocks, a block is composed of hundreds of pages, and each page consists of a data area and an Out-Of-Band (OOB) area. The OOB area is used to store information such as logical address information to identify which data is stored in the data area and Error Correction Code (ECC) to recover even if the data stored in the page is corrupted.



NAND flash memory provides three operations: read, program, and erase operation. As summarized in Table~\ref{table:nand}, each flash operation has different operation unit size and access time. In terms of operation unit size, read and program operation are performed on a page basis, while erase operation is performed on a block basis. In terms of access time, read, program, and erase operation take tens, hundreds, and thousands of $\mu$s, respectively.



NAND flash can be classified into Single-Level Cell (SLC) and Multi-Level Cell (MLC) depending on how many states are encoded in one cell. An SLC NAND flash memory can store one bit by encoding two states for each cell, while an $n$-bit MLC with $n$ greater than two can store $n$ bits by encoding $2^n$ states into one cell. Since MLC NAND can store more information with the same area than SLC NAND, 2-bit and 3-bit MLC flash memories are mainly used as storage media in recent NAND flash-based storage.



Previous planar NAND flash memories have evolved their technology by shrinking cell sizes, but have reached technical limits at 1x nm due to the extreme degradation of performance and reliability. The latest 3D NAND flash memory has overcome the structural limitations of planar NAND flash memory by vertically stacking the cell structure~\cite{3d_nand}. In addition to the multi-plane operation, 3D NAND flash memory provides a one-shot program operation~\cite{oneshot} that shortens the average effective program access time per page by programming data of pages sharing the same word line at a time.



While there are many advantages, such as fast random access, small size, and shock resistance, NAND flash memory has the following constraints: sequential program, erase-before-program, and limited endurance constraint. First, sequential program constraint indicates that pages must be programmed in ascending order within a flash block. Second, the erase-before-program constraint means that the erase operation must precede the program operation in order to recycle a programmed block. Finally, the limited endurance constraint is that NAND manufacture will no longer guarantee reliable data access from/to a flash block once the number of program/erase cycles exceeds the number listed on the datasheet.



\definecolor{Gray}{gray}{0.85}
\definecolor{LightCyan}{rgb}{0.88,1,1}

\newcolumntype{a}{>{\columncolor{Gray}}c}
\newcolumntype{b}{>{\columncolor{white}}c}

\begin{scriptsize}
\begin{table}[!t]
\renewcommand{\arraystretch}{1.2}
\small
\centering
\begin{tabular}{|c|c|c|c|c|}
\hline \hline
\multicolumn{3}{|a|}{2-bit MLC 128Gb Flash} & \multicolumn{1}{a|}{V1} & \multicolumn{1}{a|}{V2} \\ \hline \hline
 \cellcolor{Gray} & \cellcolor{Gray} & \multicolumn{1}{a|}{Data} & 8K & 16K \\ \hhline{|>{\arrayrulecolor{Gray}}->{\arrayrulecolor{black}}|>{\arrayrulecolor{Gray}}->{\arrayrulecolor{black}}|---}
                 \cellcolor{Gray} & \multirow{-2}{*}{\cellcolor{Gray}Page (Bytes)} & \multicolumn{1}{a|}{OOB} & 896 & 1.5K \\ \hhline{|>{\arrayrulecolor{Gray}}->{\arrayrulecolor{black}}|----}
\multirow{-3}{*}{\cellcolor{Gray}Unit Size} & \multicolumn{2}{a|}{Block (Bytes)} & 3M + 336K & 4M + 384K \\ \hline
                 \cellcolor{Gray} & \multicolumn{2}{a|}{Page Read ($\mu$s)} & 49 & 35 \\ \hhline{|>{\arrayrulecolor{Gray}}->{\arrayrulecolor{black}}|----}
                 \cellcolor{Gray} & \multicolumn{2}{a|}{Page Program ($\mu$s)} & 600 & 390 \\ \hhline{|>{\arrayrulecolor{Gray}}->{\arrayrulecolor{black}}|----}
                 \cellcolor{Gray} & \multicolumn{2}{a|}{Block Erase (ms)} & 4 & 4 \\ \hhline{|>{\arrayrulecolor{Gray}}->{\arrayrulecolor{black}}|----}
\multirow{-4}{*}{\cellcolor{Gray}Access Time} & \multicolumn{2}{a|}{Data Transfer (MB/s)} & 533 & 667 \\ \hline \hline
\end{tabular}
\caption{Specification of 2-bit 3D NAND flash memory.}
\label{table:nand}
\end{table}
\end{scriptsize}

\subsection{Solid State Drive}

Solid State Drive (SSD) is a storage device that uses NAND flash memory as a storage medium, unlike Hard Disk Drive (HDD) that uses disk as a storage medium. SSDs manage NAND flash memory chips using multi-channel, multi-way flash controller~\cite{ozone, multiway}.
Each channel can transfer data in parallel, only one chip can transfer data at the same time within one channel since the flash chips connected to the channel share a data bus. In order to drop-in replace HDDs that have already dominated the storage system, SSDs use a block device interface compatible with the HDDs. The host system considers a storage device that supports a block interface as a set of logical blocks or sectors, and the host can read or write successive sectors from any logical block address (LBA) of the storage device. The host system can send multiple outstanding read and write commands to fully utilize the storage performance. Block device interface includes ATA~\cite{ata} and SCSI~\cite{scsi} which assume HDDs as storage devices and NVM express (NVMe) which assumes SSDs as storage devices.

Since SSDs use NAND flash memory that supports read, program, and erase operation as a storage medium, they adopt a software called Flash Translation Layer (FTL) to be compatible with the block device interface using read and write command. FTL converts the read and write commands of the block device interface into flash read, program, and erase operations by programming the data of the write commands from the host to the unwritten flash pages and internally maintaining the logical-to-physical address mapping table. According to mapping granularity, FTL is classified as block-level mapping FTL~\cite{blockftl}, page-level mapping FTL~\cite{dftl, cdftl, tpftl}, hybrid mapping FTL~\cite{bast, fast, superblock}. Most SSDs adopt the page-level mapping FTL scheme due to its superior random write performance. SSDs are equipped with several low-power embedded processors such as ARM and ARC to execute FTL.


SSDs have GB-sized DRAM, which stores FTL metadata such as logical-to-physical mapping table and caches host data to be accessed in the near future. For page-level mapping FTL, most DRAM space is used to store mapping table. The mapping table size of the page-level mapping FTL increases in proportion to the capacity of the SSD. For example, a 1TB SSD requires 1GB DRAM to store the logical-to-physical mapping table. If the DRAM capacity is insufficient to load the entire mapping table, the FTL might first read the mapping table stored in the NAND flash to consult the physical address where the requested data is located. In this case, SSDs exhibit low performance thus most high-performance SSDs place the entire mapping table in DRAM when powered on. However, it is no longer affordable to adopt a DRAM proportional to the SSD capacity. Recently, there has been an increasing demand for DRAM-less SSDs equipped with the map cache unit that caches entries of mapping table in a limited amount of SRAM~\cite{dramless}.


\section{SSD Performance Exploration} \label{sec:motivation}

In this section, we analyze the SSD performance to predict which component, NAND flash memory, host interface, and FTL, will be the bottleneck of future SSD performance. We explore the performance by modeling and simulating SSD performance, which is based on Microsoft Research SSD extension~\cite{ssdext} for the DiskSim simulation environment~\cite{disksim}.


\subsection{SSD Performance Modeling} \label{sec:model}

\begin{scriptsize}
\begin{table}[!t]
\renewcommand{\arraystretch}{1.2}
\small
\centering
\begin{tabular}{|c|l|}
\hline \hline
\multicolumn{1}{|a|}{Symbol} & \multicolumn{1}{a|}{Description} \\ \hline \hline
\multicolumn{1}{|a|}{$T_{FTL\_map}$} & The FTL execution time of processing map lookup. \\ \hline
\multicolumn{1}{|a|}{$T_{FTL\_cmd}$} & The FTL execution time of issuing NAND request. \\ \hline
\multicolumn{1}{|a|}{$T_{NAND\_read}$} & The latency of reading data from the NAND page. \\ \hline
\multicolumn{1}{|a|}{$T_{NAND\_program}$} & The latency of programming data to the NAND page. \\ \hline
\multicolumn{1}{|a|}{$T_{NAND\_erase}$} & The latency of erasing data of the NAND block. \\ \hline
\multicolumn{1}{|a|}{$T_{NAND\_bus}$} & The latency of transferring data from/to the NAND. \\ \hline
\multicolumn{1}{|a|}{$T_{Host\_transfer}$} & The latency of transferring data from/to the SSD. \\ \hline \hline
\end{tabular}
\caption{Timing parameters for processing 4KB random read command.}
\label{table:timeline}
\end{table}
\end{scriptsize}

\begin{figure}[!t]
\begin{center}
	\vspace{-0.1cm}
	\includegraphics[width=1\columnwidth, keepaspectratio=true]{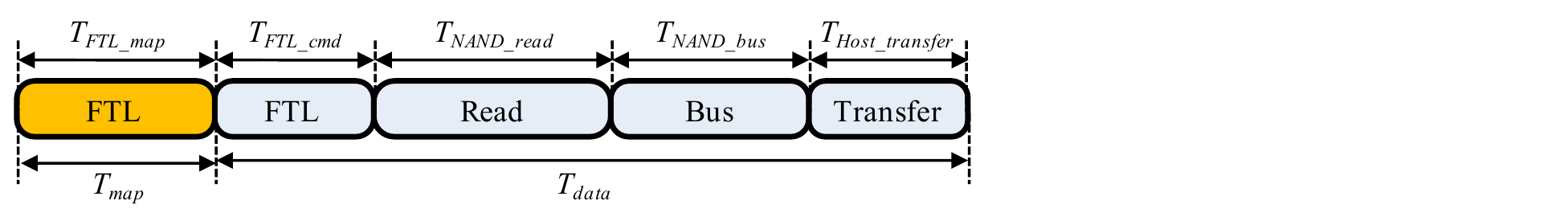}
	\vspace{0.2cm}
	(a) map cache hit
	\includegraphics[width=1\columnwidth, keepaspectratio=true]{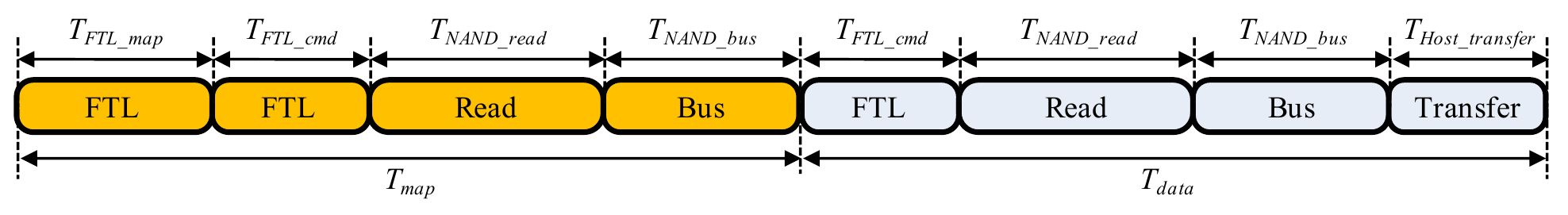}
	(b) map cache miss
	\caption{SSD timelines for processing a single 4KB random read command.} \label{fig:timeline}
\end{center}
\vspace{-0.2cm}
\end{figure}

Table~\ref{table:timeline} and Figure~\ref{fig:timeline} present timing parameters for SSD performance modeling and timeline diagrams based on the timing parameters when a SSD processes a single 4KB random read command, respectively. When receiving a read command from the host, the FTL first looks up the map cache to identify where the requested data is stored in the flash memory ($T_{FTL\_map}$). As shown in Figure~\ref{fig:timeline}(a), if there is a corresponding logical-to-physical mapping entry in the map cache, FTL issues a flash read operation to retrieve the host data from the flash memory ($T_{FTL\_cmd}$). After reading the data from the page ($T_{NAND\_read}$), the data is sent to the SSD's internal RAM ($T_{NAND\_bus}$), then the requested data is transferred to the host via host interface ($T_{Host\_transfer}$). If the corresponding mapping entry is not cached, as depicted in Figure~\ref{fig:timeline}(b), FTL issues a read operation to the NAND flash memory where the mapping table is stored and reads and loads the logical-to-physical mapping entry from the flash memory into the map cache.


To improve the 4KB random read performance, it is necessary to optimize the timing parameters included in Table~\ref{table:timeline}. First, $T_{NAND\_read}$ can be relatively shorten by increasing the number of NAND chips, since each flash chip can perform its own flash operation in parallel. Similarly, $T_{NAND\_bus}$ can be relatively reduced by increasing the number of flash channels because each channel can transfer data from/to flash memories in parallel. Next, $T_{Host\_transfer}$ can be improved by enhancing the bandwidth of the host interface, such as changing the host interface from PCI express 3.0 (PCIe) x8 to PCIe 3.0 x16. Finally, $T_{FTL\_map}$ and $T_{FTL\_cmd}$ can be reduced by increasing the frequencies of embedded processors running FTL or by increasing the number of processors.


The probability of $T_{map}$ depends on the hit ratio of the map cache, and the hit ratio is determined by the cache size and the map cache algorithm~\cite{lrfu, cflru, adaptive}. This paper does not discuss the map cache hit ratio.


\subsection{SSD Performance Analysis}

Based on the performance modeling of Section~\ref{sec:model}, we analyze and predict the 4KB random read performance of SSD. Using the timing parameters of the V2 2-bit 3D NAND flash memory as described in Table~\ref{table:nand}, we experimented with varying the number of channels and ways from 1-channel, 1-way to 16-channel, 8-way. The host interface we used is NVMe over PCIe 3.0 x16, which provides the bandwidth up to 15.76GB/s. The number of outstanding read requests is set to 512 to maximize the parallelism of NAND flash memories. $T_{FTL\_cmd}$ is changed from 0$\mu$s to 4$\mu$s and $T_{FTL\_cmd} $ is assumed to be 0$\mu$s because generating and issuing flash operations can be processed immediately.

\begin{figure}[!t]
\vspace*{-0.8cm}
\begin{center}
\hspace*{-0.5cm} \vspace*{-0.6cm}
	\includegraphics[width=1\columnwidth, keepaspectratio=true]{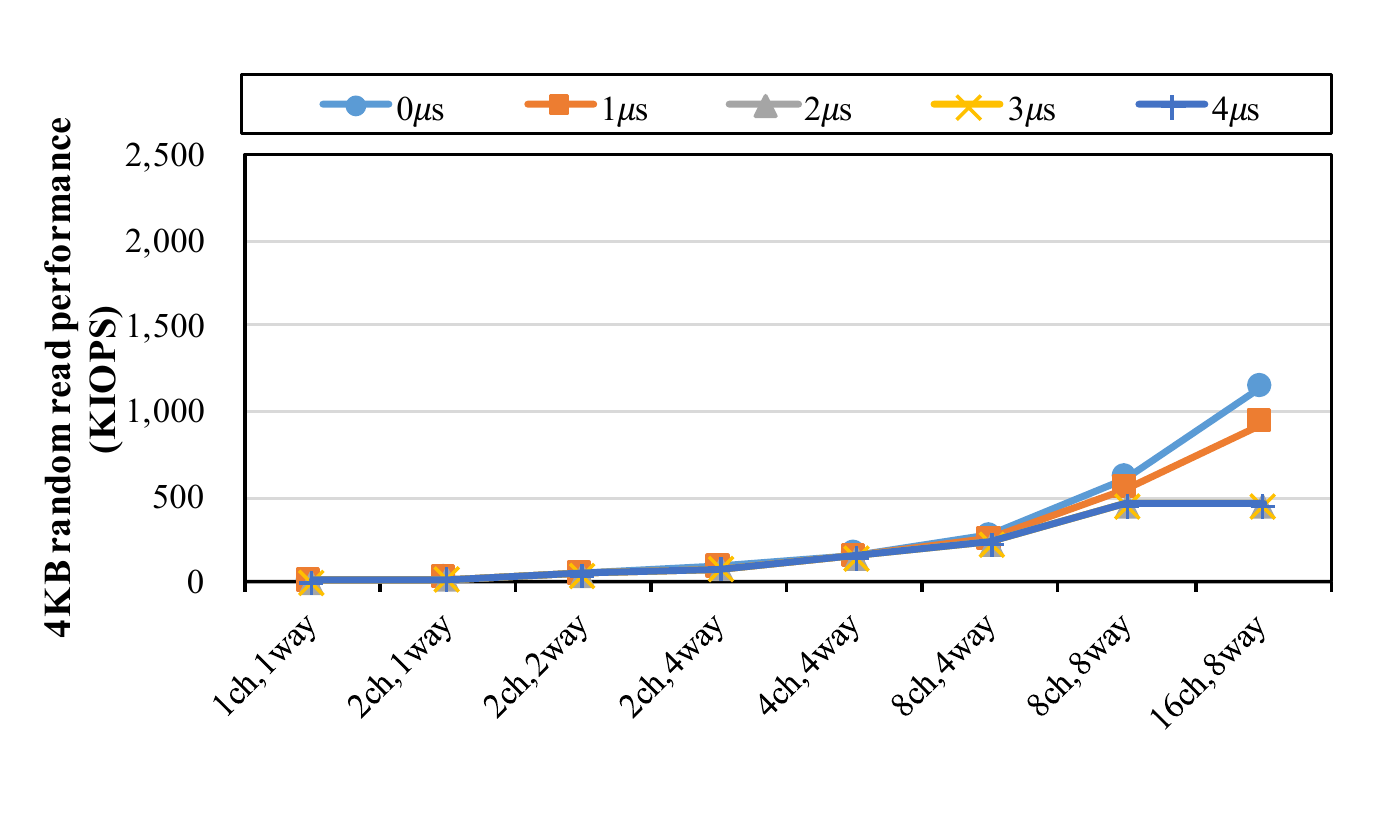}
	\hspace*{0.75cm} (b) map hit ratio = 0\%
	\caption{4KB random read performance on various configurations.} \label{fig:4kperf}
\end{center}
\vspace*{-0.3cm}
\end{figure}

Figure~\ref{fig:4kperf}(a) and Figure~\ref{fig:4kperf}(b) illustrate the SSD performance of 4KB random read with map cache hit and map cache miss, respectively. When the map cache hit ratio is set to 100\%, the FTL execution time becomes the bottleneck of 8-channel, 8-way SSD where FTL execution time is 1$\mu$s to process a single 4KB read command. If the map cache hit ratio is set to 0\%, $T_{NAND\_read}$ and $T_{NAND\_bus}$ are executed twice for each command, and this allows for FTL with 1$\mu$s execution time to be the bottleneck of SSD performance starting from 16-channel, 8-way. As a result, FTL execution time later becomes the SSD performance bottleneck compared to the map cache hit case.


\begin{figure}[!t]
\vspace*{-0.5cm}
\begin{center}
\hspace*{-0.5cm} \vspace*{-0.6cm}
	\includegraphics[width=1\columnwidth, keepaspectratio=true]{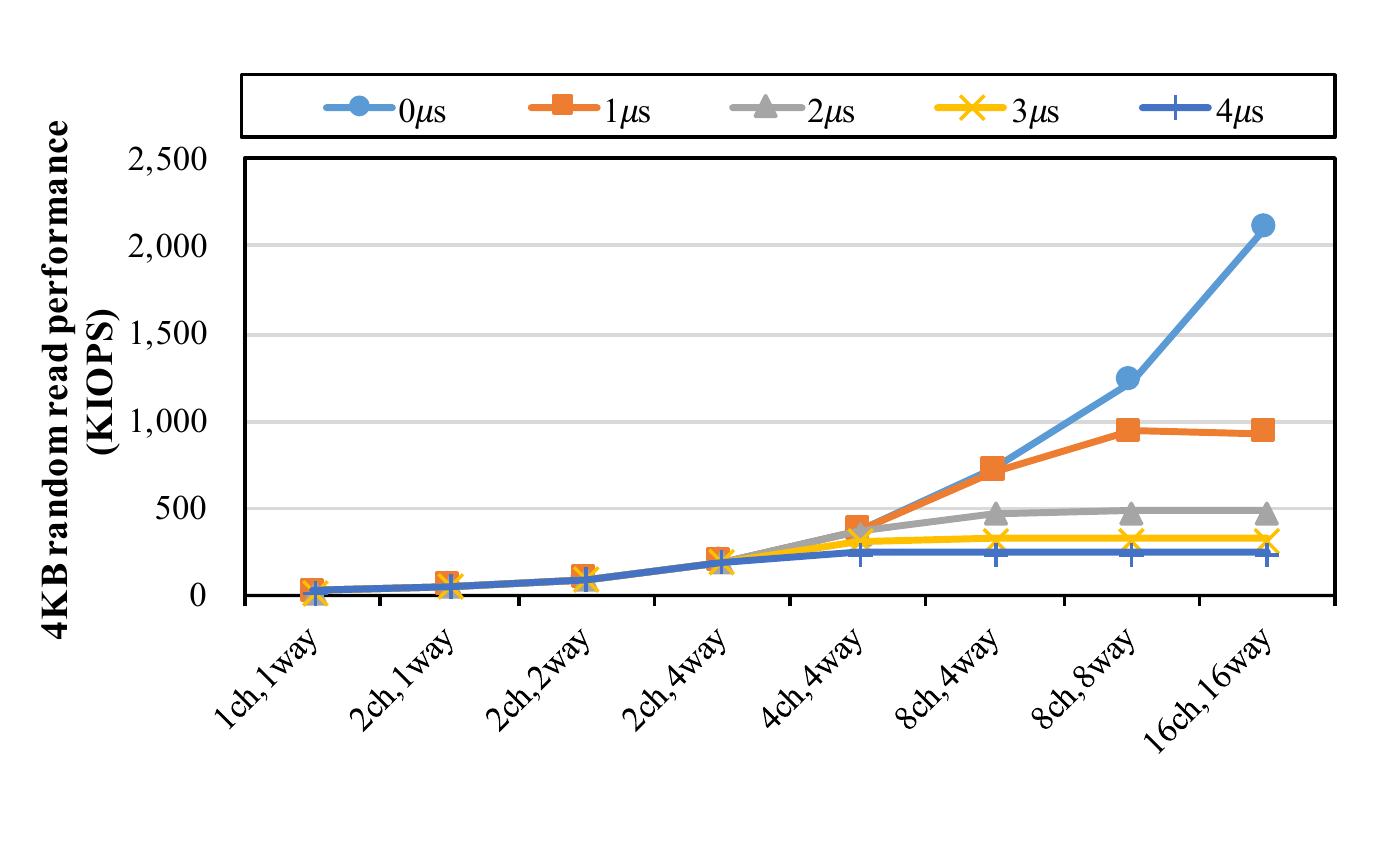}
	\hspace*{0.75cm} (a) map hit ratio = 100\%
\end{center}
\end{figure}

\begin{figure}[!t]
\vspace*{-0.5cm}
\begin{center}
\hspace*{-0.2cm} \vspace*{-0.5cm}
	\includegraphics[width=1.1\columnwidth, keepaspectratio=true]{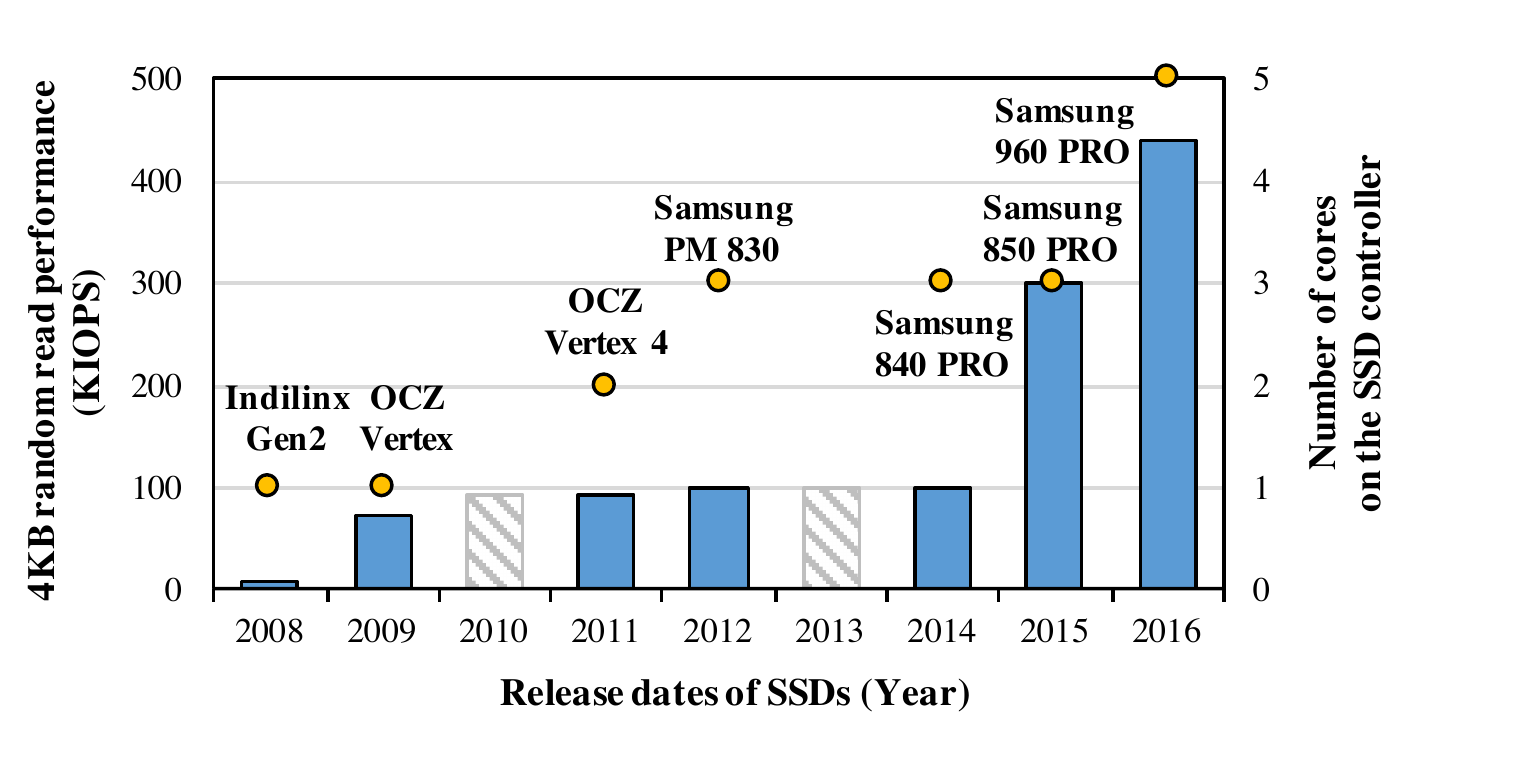}
	\caption{Changes in the number of cores equipped with major SSD controllers from 2008 to 2016~\cite{barefoot2, vertex, vertex4, 830, 840, 850, 960}.} \label{fig:samsungssd}
\end{center}
\end{figure}

To prevent the FTL execution time from becoming a bottleneck in SSD performance, there are two simple approaches: increasing the clock frequencies of the processors that drive the FTL, or increasing the number of processors. Since scaling frequency leads to excessive processor power consumption, existing SSDs have increased the number of CPU cores so that FTL execution time does not become bottleneck in SSD performance, as shown in Figure~\ref{fig:samsungssd}. In this approach, the entire mapping table is statically partitioned and allocated to the FTL instance operating in different CPU cores to avoid synchronization overhead among FTL instances. Each FTL instance handles the read and write commands for the logical address space covered by its mapping table. The data structures, however, such as the map of mapping table can be accessed simultaneously by the FTL instances running on different cores, and this overhead increases as the number of CPUs increases. Increasing the number of CPUs also increases the price and power consumption of SSDs, so increasing the number of CPUs is no longer a scalable solution.


We propose a Flash Map Management Unit (FMMU) to prevent FTL execution time from becoming an SSD performance bottleneck as SSD performance improves. FMMU is a single hardware-automated map cache unit that can handle requests much faster than the FTL map cache running on the embedded processors. Since FMMU is a single unit, there is no synchronization overhead. The following section introduces the design of FMMU.


\section{Design of FMMU} \label{sec:design}

FMMU is a hardware-automated unit that manages and caches logical-to-physical address mapping tables. FMMU is designed to reduce the number of flash memory operations while simultaneously maximizing the parallelism of flash memory by processing concurrent map requests in a non-blocking manner. FMMU ensures that FTL execution time does not become a bottleneck in SSD performance even if SSD performance is further improved in the future. Since FMMU supports packetized interface, FMMU can easily substitute existing map cache unit implemented by software. Section~\ref{sec:overview} introduces the generic FTL framework and briefly explains how to use FMMU in the framework. Sections~\ref{sec:fmmu} to \ref{sec:arb} describe the design of FMMU in detail.



\subsection{FTL Framework} \label{sec:overview}

\begin{figure}[!t]
\begin{center}
	\includegraphics[width=0.9\columnwidth, keepaspectratio=true]{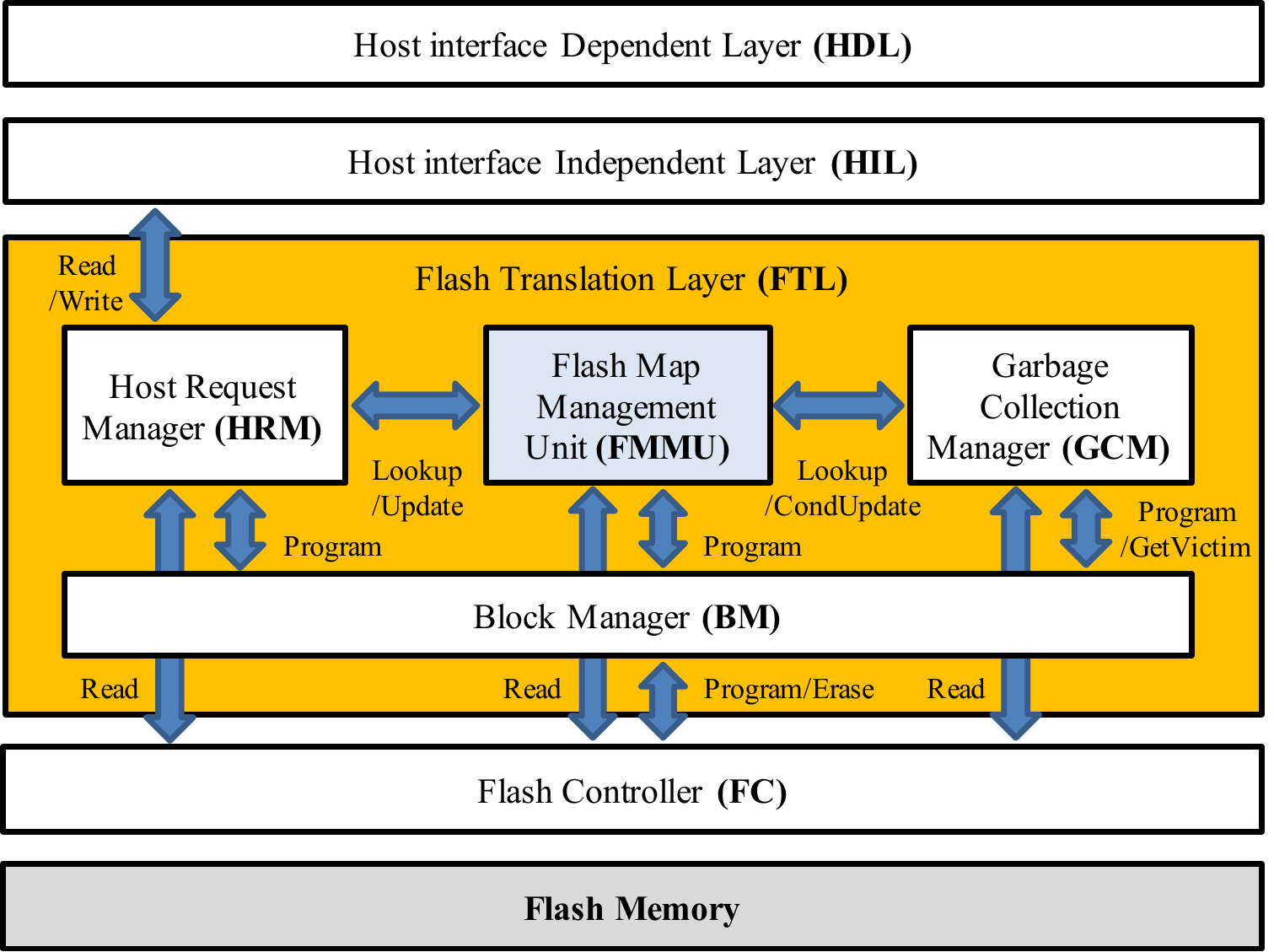}
	\caption{Architecture of NAND flash-based storage system.} \label{fig:framework}
\vspace*{-0.3cm}
\end{center}
\end{figure}

Figure~\ref{fig:framework} shows the architecture of a typical NAND-flash based storage system. Read and write commands from the host are passed to the FTL through the Host interface Dependent Layer (HDL) and Host interface Independent Layer (HIL). FTL converts read and write requests to flash read, program, and erase operations and issues them to the Flash Controller (FC).



\noindent\textbf{Host interface Dependent Layer (HDL).} Host systems access NAND-flash based storage by various block device interfaces such as ATA, SCSI, eMMC, and NVMe. HDL communicates with the host by exchanging command, response, and data through the protocol defined by each host interface. The HDL processes host interface specific commands and sends read and write requests to the HIL, that are required for all storage regardless of the host interface.


\noindent\textbf{Host interface Independent Layer (HIL).} The HIL passes the read and write requests to the FTL after performing two tasks: the request chopper and the dependency checker. The request chopper splits a request into several sub-requests with a predefined size. A host can read or write multiple consecutive sectors from a device with a single command. For example, a command in the NVMe protocol can read or write a maximum of 65,536 sectors, which is up to 256MB when using a 4KB sector. FTL stores and manages the data of requests and the status of ongoing requests in RAM, and the RAM usage increases in proportion to the request size. By splitting the request and passing the sub-request to the FTL, the request chopper allows the FTL to handle large commands while using limited RAM usage.



The dependency checker identifies and resolves the data dependency between read and write commands. The host system sends several concurrent commands to storage, allowing storage to internally process commands out-of-order to improve performance. In this case, the result of concurrent commands processed must be the same result as if commands were sequentially executed. If the address ranges of the read and write commands overlap each other, a data dependency occurs. If the dependency checker finds such a data dependency, it rearranges or pends the requests to make the FTL data dependency-oblivious.



\noindent\textbf{Flash Translation Layer (FTL).} The FTL consists of four basic modules: the Host Request Manager (HRM), the Garbage Collection Manager (GCM), the Block Manager (BM), and the Flash Map Management Unit (FMMU). HRM handles read and write requests received from HIL using FMMU and BM. HRM processes the read request by sending a \textit{Lookup} request to the FMMU to figure out the physical address corresponding to the logical address of the request, and then sending a flash read request to the FC for the physical address. On the other hand, HRM first writes data to the unwritten flash page by sending a program request to the BM, and then updates the logical-to-physical mapping by sending a \textit{Update} request to the FMMU for the write request.



BM converts program requests requested by HRM, GCM, and FMMU into flash program requests and manages flash blocks. In order to maximize the write performance of the SSD, BM collects the program requests and sends a flash program request of the multi-plane and one-shot program operation to the FC. It also sends a number of outstanding flash program requests to FC to maximize the parallelism of the multi-channel and multi-way flash controller. The BM manages the flash blocks according to the number of valid pages, selects and erases blocks that do not have a valid page if free blocks are needed, and picks out the block with the least valid page if garbage collection is required.



GCM performs garbage collection, allowing the BM to have enough free blocks. If the BM provides a victim block to the GCM, the GCM first reads all the flash pages of the victim block and identifies the logical addresses recorded in the OOB area of each page. It then determines whether each page is valid by comparing the physical address of the page with the physical address mapped to the logical address obtained from the \textit{Lookup} request of FMMU. The GCM sends program requests to the BM for each valid page and updates the mapping table via \textit{CondUpdate} of the FMMU when the program operation is complete.



FMMU manages the logical-to-physical mapping table and processes \textit{Lookup}, \textit{Update}, and \textit{CondUpdate} requests generated by HRM and GCM. FMMU stores the entire mapping table in flash memory and caches mapping entries according to RAM capacity. When a map cache miss occurs during processing of \textit{Lookup} and \textit{Update} request, the FMMU sends a flash read request to the FC to load the mapping table stored in flash. If the number of dirty cache blocks in the FMMU exceeds a certain threshold, a program request is sent to the BM to make a clean cache block and the dirty map table is written to flash. For GCM, FMMU provides a special interface \textit{Condupdate}. While GCM is copying the valid page, the HRM can program the latest version of the data for the same logical address of the page and concurrently issue the \textit{Update} request. If a valid page copy is completed later and the GCM requests a map update, the FMMU should not perform this mapping table update. The \textit{CondUpdate} request has old physical address information in addition to the logical address and physical address. For the \textit{CondUpdate} request, FMMU ensures correctness by performing map update only if the latest map table still points to the old physical address.

\subsection{Flash Map Management Unit} \label{sec:fmmu}

\begin{figure}[!t]
\begin{center}
	\hspace{-0.4cm}
	\includegraphics[width=1.03\columnwidth, keepaspectratio=true]{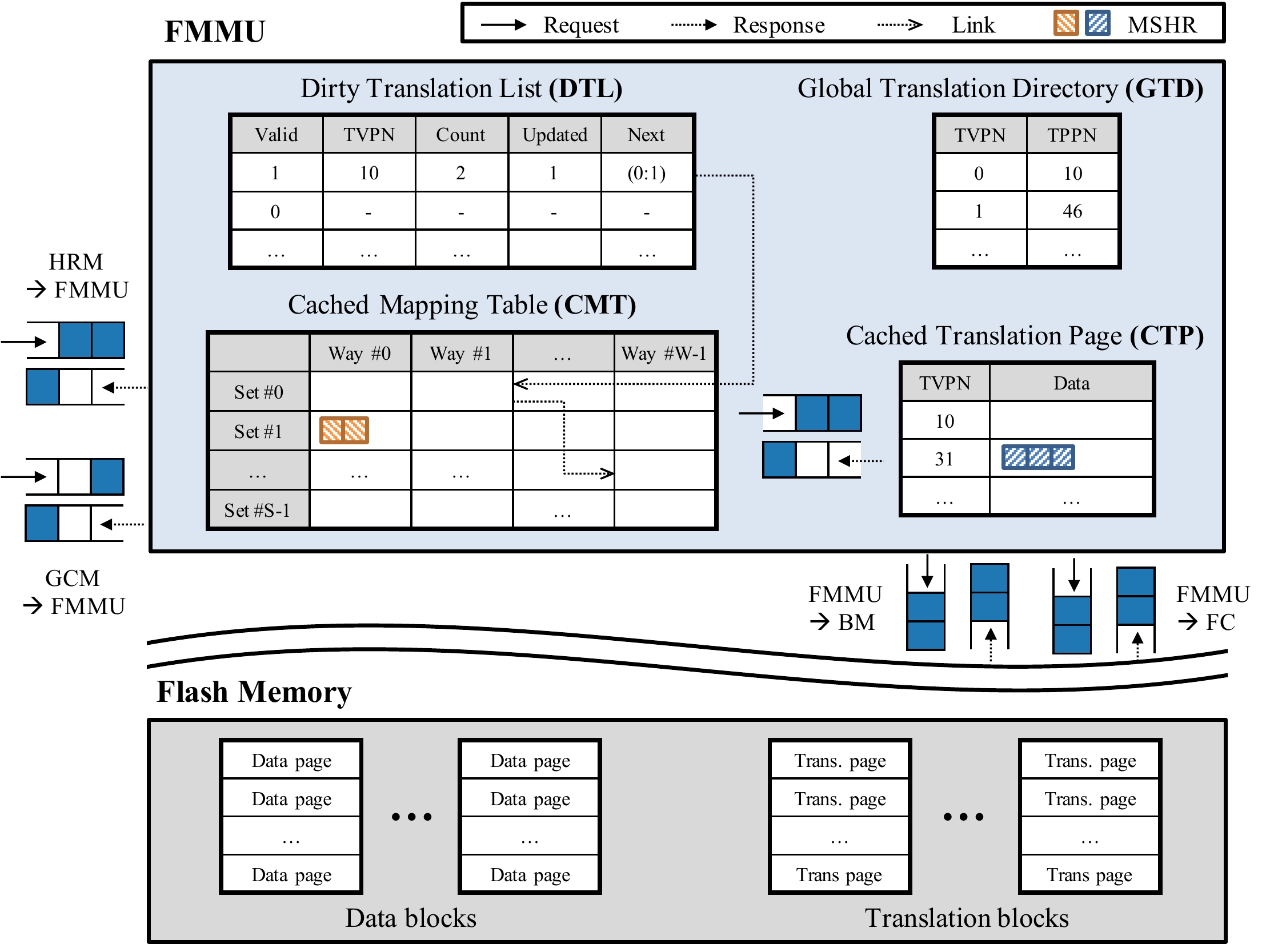}
	\vspace{-0.5cm}
	\caption{Architecture of FMMU.} \label{fig:fmmu}
\vspace*{-0.2cm}
\end{center}
\end{figure}

As shown in Figure~\ref{fig:fmmu}, FMMU adopts a demand-based page-level address mapping scheme and partitions the flash memory into data blocks storing the host data and translation blocks storing the page-level mapping table. One data page stores data of consecutive logical block addresses. A logical-to-physical mapping for a data page is represented by a Data Logical Page Number (DLPN)-to-Data Physical Page Number (DPPN). Similarly, one translation page stores consecutive DLPN-to-DPPN entries, and the translation page mapping is represented by the Translation Virtual Page Number (TPPN)-to-Translation Physical Page Number (TPPN). When the SSD is powered up, the FMMU populates all TVPN-TPPN entries into the global translation directory (GTD) in the RAM area.



FMMU adopts a two-level caching mechanism and caches DLPN-to-DPPN entries in Cached Mapping Table (CMT) and Cached Translation Page (CTP), which are hardware state machines that operate concurrently. The first-level cache, CMT, is designed to exploit the temporal and spatial locality of access patterns of HRM and GCM requests to reduce the number of accesses to translation pages. Each cache block of CMT stores consecutive DLPN-to-DPPN entries. If a cache miss occurs or a flush is required in the CMT, the CMT generates a load or flush request to the CTP, which is the second-level cache. The CTP is designed to reduce the number of flash operations to load or flush the translation page by exploiting the locality of the CMT access pattern. If a cache miss occurs or a flush is required by the CTP, the CTP generates and sends a flash read and a flash program request to the FC and BM, respectively.



\begin{figure}[!t]
\begin{center}
	\includegraphics[width=1\columnwidth, keepaspectratio=true]{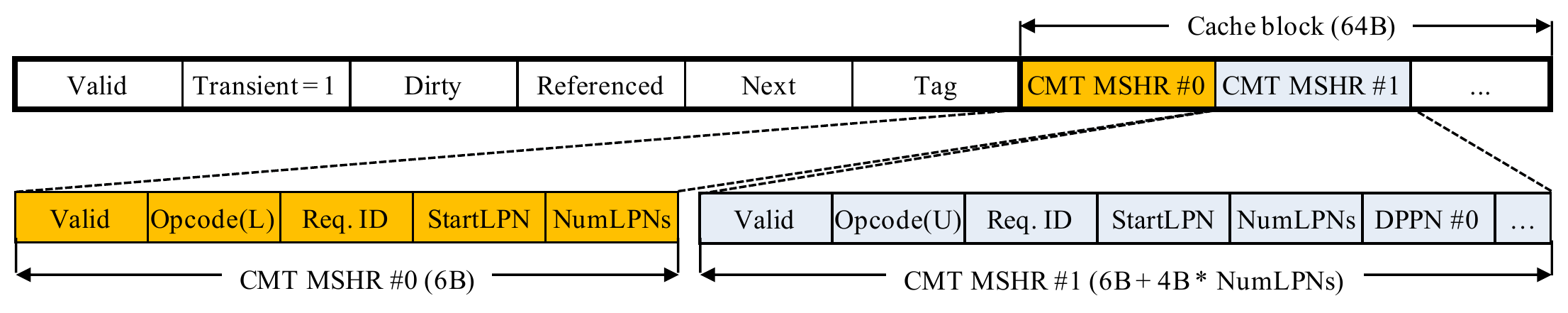}
	\vspace{0.1cm}
	(a) transient = 1
	\vspace{0.2cm}
	\includegraphics[width=0.99\columnwidth, keepaspectratio=true]{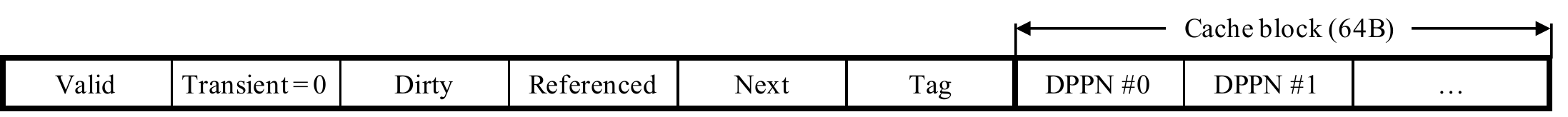}
	(b) transient = 0
	\vspace{-0.2cm}
	\caption{Cache block layout of CMT. MSHR \#0 and \#1 represent outstanding cache misses of \textit{Lookup} and \textit{Update} request, respectively.} \label{fig:cmt}
\end{center}
\end{figure}

\begin{figure}[!t]
\begin{center}
	\includegraphics[width=1\columnwidth, keepaspectratio=true]{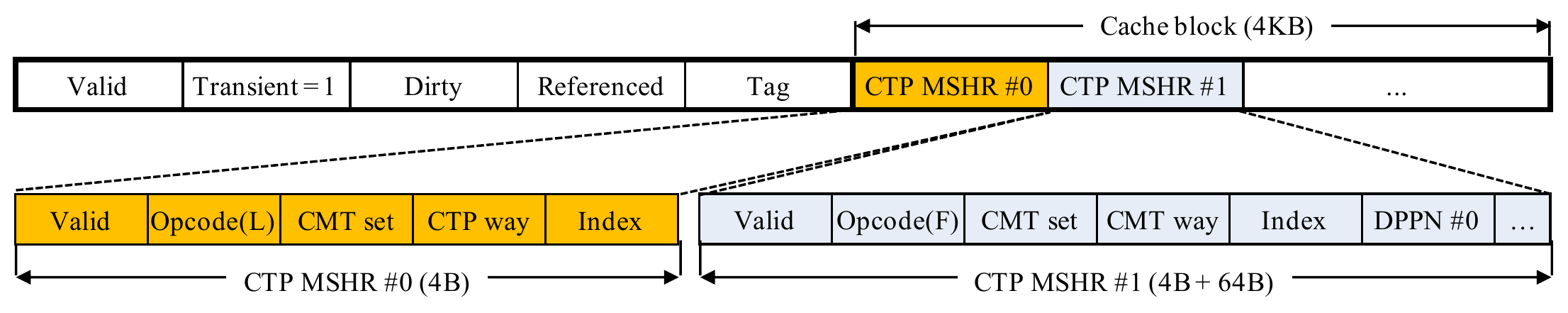}
	\vspace{0.1cm}
	(a) transient = 1
	\vspace{0.2cm}
	\includegraphics[width=0.99\columnwidth, keepaspectratio=true]{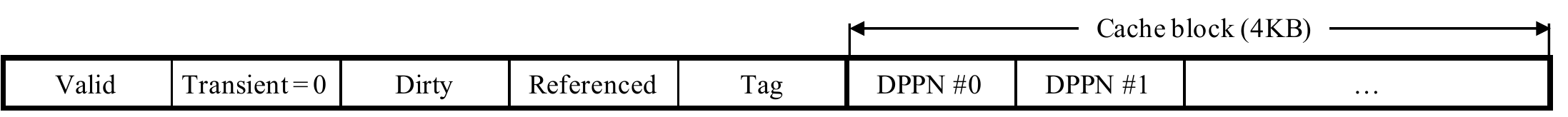}
	(b) transient = 0
	\vspace{-0.2cm}
	\caption{Cache block layout of CTP. MSHR \#0 and \#1 represent outstanding cache misses of load and flush request, respectively.} \label{fig:ctp}
\end{center}
\end{figure}

In order to deal with cache misses generated by hardware-implemented CMT and CTP in a non-blocking manner, FMMU adopts the in-cache MSHR scheme which logs the request information that caused a cache miss in the form of a Miss Status Holding Register (MSHR)~\cite{incache}. Each CMT and CTP cache block has a transient bit for each block to indicate whether MSHRs are being logged in the cache block to handle the current cache miss or not. Figure~\ref{fig:cmt} and Figure~\ref{fig:ctp} describe the cache block layout of CMT and CTP, respectively. When a cache miss occurs for the HRM or GCM request, the CMT logs the miss request information in the allocated cache block as shown in Figure~\ref{fig:cmt}(a) and sends the load request to the CTP. When a response to the load request arrives from the CTP, CMT store the DLPN-to-DPPN entries and sends a response to the HRM or GCM by referring to the MSHR information that was logged in the cache block as illustrated in Figure~\ref{fig:cmt}(b). Similarly, when a cache miss occurs for a CMT request, the CTP logs the miss request information in the CTP cache block as the MSHR format and sends a flash read request to the FC as depicted in Figure~\ref{fig:ctp}(a). When a response to a flash read request arrives, the CTP places the map data on the cache block and sends the response to the CMT as shown in Figure~\ref{fig:ctp}(b). This scheme is space efficient because it does not require additional RAM to record the information of outstanding miss requests, but uses cache block space that is not utilized. Since CMT and CTP handle concurrent HRM/GC requests and CMT requests in a non-blocking manner, FMMU can generate multiple flash requests and maximize the parallelism of flash memory.


The CMT sends a flush request to the CTP when the number of dirty cache blocks exceeds a certain threshold. In order to quickly find the cache blocks belonging to the same translation page and perform a batch update, the CMT adopts the Dirty Translation List (DTL) and tracks the next information for each cache block, as presented in Figure~\ref{fig:fmmu} and Figure~\ref{fig:cmt}, respectively. If a CMT cache block becomes dirty by \textit{Update} or \textit{CondUpdate} request, the CMT checks whether the TVPN is already registered in the DTL. If it is not registered, TVPN is registered in the tail entry of DTL and next information of the entry holds the set and way information of current cache block. If the corresponding DTL entry exists, the next information stored in the entry is saved in the next information of the current cache block and the next information of the entry is replaced with the current cache block address. When flushing CMT, dirty cache blocks corresponding to TVPN can be searched along the next link only. While flushing the dirty blocks of the CMT, the dirty cache blocks corresponding to the same TVPN can be visited by the next information.


\begin{figure*}[!t]
\begin{center}
\hspace*{0cm}
	\includegraphics[height=0.26\textheight, keepaspectratio=true]{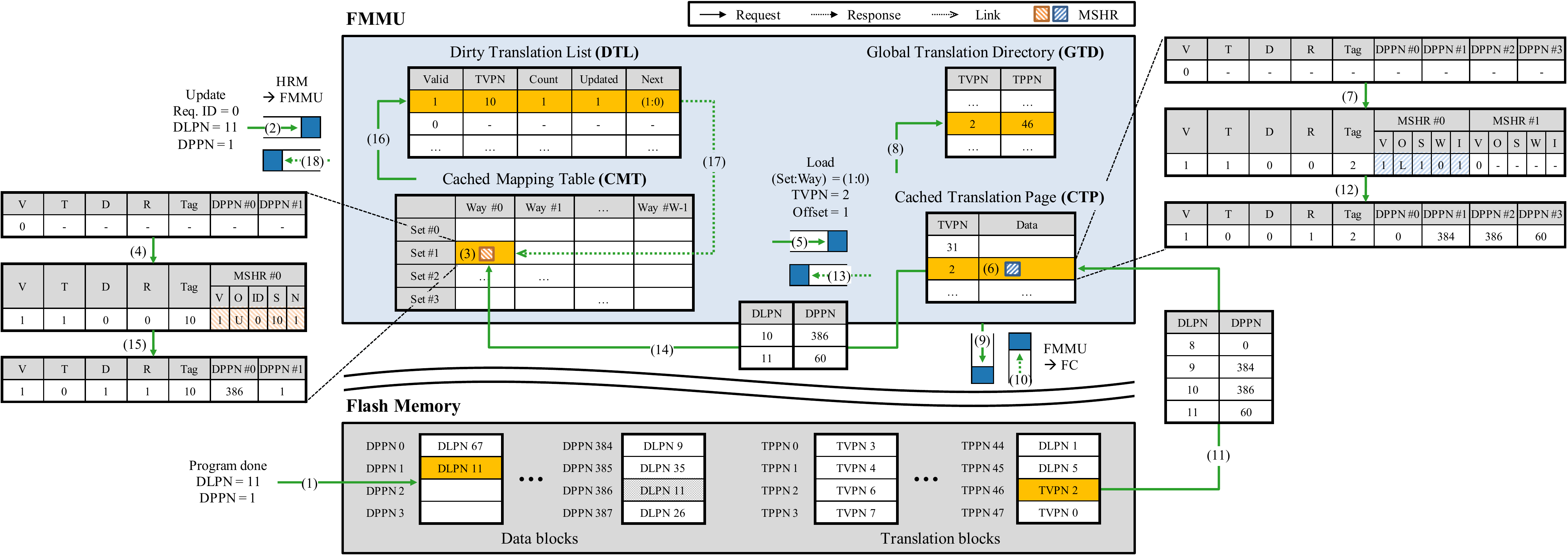}
	\caption{Suppose that the translation page stores four DPPNs, and the CMT and CTP cache blocks store two and four DPPNs, respectively. (1)-(2) HRM programs data of DLPN 10 to DPPN 1 and sends an \textit{Update} request. (3)-(5) Upon a cache miss in CMT, cache block is allocated, and request is recorded as MSHR. (3)-(5) Determine that a cache miss occurs, allocate a cache block, and record the request as an MSHR. (5) Send load request of TVPN 2 (11/4=2) and offset 1 (3/2=1) to CTP. (6)-(7) Upon a cache miss in CTP, and request is recorded as MSHR after cache block allocation. (8)-(9) Look up GTD and send a flash read request to TPPN 46 for TVPN 2. (10)-(13) Store the map data of TVPN 2 in the cache block and send the response to the CMT. (14)-(17) Save the map data in the cache block, and since TVPN 2 is not in the DTL, register the cache block in the new DTL entry. (18) Finally, send response to HRM of the \textit{Update} request.} \label{fig:address}
	

\end{center}
\end{figure*}

\subsection{Address Translation Process}

When the HRM or GCM sends a request to the FMMU, the FMMU processes the address translation through the CMT and the CTP. First, the CMT obtains the set and tag values ​​from the DLPN of the request and finds the cache block whose tag value matches among the cache blocks in the set. If a cache miss occurs, the CMT selects and evicts the victim cache block by the replacement policy, allocates the block for DLPN, and logs the MSHR information for the request. It then sends a load request to the CTP, the second-level cache, for the TVPN containing the mapping entry of the DLPN. Similar to CMT, CTP also determines cache hit/miss. If cache miss occurs, it selects victim block, and then allocates the block for TVPN and logs MSHR for load request of CMT. To load the translation page, the CTP looks up the GTD to find out the TPPN for the TVPN and sends a flash read request to the FC for the corresponding TPPN. When a response to a read request is received from the FC, the CTP stores the map data in the cache block and sends a response by referring to the MSHR that has been logged by the CMT. When the CMT receives a response from the CTP, it finally stores the map data in the cache block and sends a response to the HRM or GCM based on the logged MSHR. Figure ~\ref{fig:address} shows an example of the address translation process of the FMMU when a cache miss occurs in CMT and CTP for a \textit{Update} request.

\subsection{Replacement Policy}

When a CMT or CTP encounters a cache miss while processing a request, it performs a replacement process that evicts and allocates a cache block for a given request. Both CMT and CTP select a victim cache block based on least recently used (LRU) algorithm among non-dirty cache blocks because evicting a dirty cache block involves long latency operations. We apply a second-chance algorithm that maintains a referenced bit for every cache block that can be implemented in hardware. If the states of all cache blocks in the specified set are dirty, the request is not processed until the non-dirty cache block is generated by the flush request.


\subsection{Flush Policy}

CMT and CTP try to keep the number of non-dirty cache blocks above the threshold through flush operation because requests can be blocked if the non-dirty cache block is insufficient. When the number of non-dirty cache blocks reaches the low watermark, CMT and CTP alternate between flushing a certain number of cache blocks and processing a request until the the number reaches the high watermark. This prevents the request from experiencing long latency by staying in the request queue for a long time by flush operation.


CMT chooses flush victim by cost-benefit analysis for three criteria: greedy, oldest-update-first, and least-recently-updated. Greedy selects the TVPN with the most dirty cache blocks in CMT. Oldest-update-first selects the TVPN of the entry that was least-recently-registered in the DTL. Finally, least-recently-updated selects the TVPN that was the least-recently-updated by \textit{Update} or \textit{CondUpdate} in CMT. To determine these criteria, DTL maintains head and tail pointers and registers new entries from the tail. All entries record the number of dirty cache blocks corresponding to the TVPN and manage the updated bit to see if it has recently been updated. Unlike CMT, CTP tracks TVPNs in the order of flush requests issued by CMT, and flushes TVPNs in the same order if the number of dirty cache blocks exceeds threshold.


\subsection{Arbitration Mechanism} \label{sec:arb}

\begin{figure}[!t]
\begin{center}
	\includegraphics[width=1\columnwidth, keepaspectratio=true]{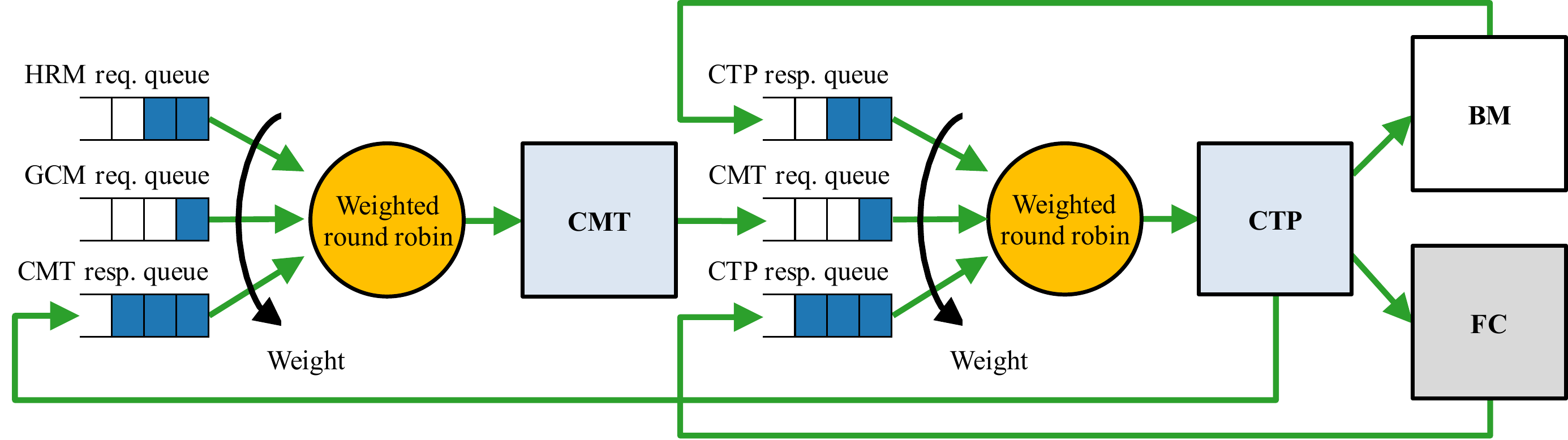}
	\caption{Arbitration mechanisms of CMT and CTP.} \label{fig:arbitration}
\end{center}
\end{figure}

CMT and CTP use individual request / response queues for each module that communicates with each other independently so that queues do not need to be locked. CMT and CTP reactively select a queue to fetch a request / response packet, and if the queue is not empty, fetch the packet and process it. As illustrated in Figure~\ref{fig:arbitration}, the queue to be fetched is chosen by the weighted round robin manner. Response queues have a higher weight than request queues because they can send responses to multiple pending requests in batches by processing one response packet. The weight of the HRM and GCM request queues of the CMT are changed at runtime by the valid page count of the GC victim block to minimize the fluctuation of SSD write performance.



\section{Evaluation} \label{sec:evaluation}

\subsection{Experimental Setup}

We implemented the SSD simulator based on Microsoft Research SSD extension~\cite{ssdext} for the DiskSim simulation environment~\cite{disksim} to evaluate the FMMU. The SSD simulator models a 16GB capacity with 16-channel, 8-way NAND configuration and allocates 15\% of the total capacity as over-provisioning space. Each NAND flash chip has two planes, and the timing parameter of the flash memory is set to V2 2-bit 3D NAND flash memory as show in Table~\ref{table:nand}. The SSD simulator reads data from flash with multi-plane read operations and programs data to flash with multi-plane and one-shot programming operations. The SSD simulator uses NVMe over PCIe 3.0 x16, which provides about 15.76 GB of bandwidth, as the host interface.


The FMMU is evaluated by comparing with the prior map cache schemes DFTL and CDFTL. For a fair comparison, the SSD simulator uses the general FTL framework and adopts DFTL, CDFTL and FMMU as map cache units. The SSD simulator uses 1,088KB of RAM for the map cache unit. DFTL allocates 1,088KB of RAM for CMT, and both CDFTL and FMMU allocate 64KB and 1,024KB of RAM for CMT and CTP, respectively. All map cache schemes use the second chance algorithm as a replacement policy.


The SSD simulator also models the execution time of FTLs. Execution times of DFTL and CDFTL are obtained by implementing and executing DFTL and CDFTL in Gem5 Simulator~\cite {gem5}. The CPU model used in the Gem5 simulator is the 400MHz ARM Cortex-R4, which is the same CPU that is employed in recent SSDs. Unlike DFTL and CDFTL, FMMU should be implemented as hardware, so the execution time of FMMU is approximated using Xilinx High-Level Synthesis (HLS)~\cite{hls}. For a fair comparison with the Cortex-R4 running at 400MHz, the clock speed in the HLS is set to 2.5ns.


We evaluated the SSD using synthetic workload and block I/O traces. Synthetic workloads include 4KB random read/write and 64KB sequential read/write. Random workload and sequential workload performance are measured by I/O Operations Per Second (IOPS) and bandwidth (GB/s), respectively. To understand the performance in-depth, the utilizations of the following SSD components are measured based on the busy time during the evaluation: the host interface, the NAND bus, the NAND chip, and the FTL. The block I/O traces used in the experiments are MSR\_proj~\cite{msr}, MSR\_hm~\cite{msr}, WebSearch~\cite{web} and the characteristics of the traces are summarized in Table~\ref{table:trace}. MSR\_proj and MSR\_hm are write dominant 1-week block workloads collected from Microsoft Research Cambridge's enterprise servers. MSR\_proj and MSR\_hm are server traces for project directory and hardware monitoring, respectively. WebSearch is a read-dominant workload collected by the Storage Performance Council. The performance is evaluated by the time the SSD simulator performed to the end of the trace. In the performance evaluation, the number of outstanding requests and the inter-arrival time are set to 512 and 0$\mu$s to fully evaluate the performance of SSDs.


Finally, the performance of the SSD with FMMU is measured by varying the configuration of the host interface and the NAND flash memory in order to evaluate whether the FMMU allows scalable performance of the SSD.


\begin{scriptsize}
\begin{table}[!t]
\renewcommand{\arraystretch}{1.2}
\small
\centering
\begin{tabular}{|c|c|c|c|c|}
\hline \hline
\multicolumn{2}{|a|}{Traces} & \multicolumn{1}{a|}{MSR\_proj} & \multicolumn{1}{a|}{MSR\_hm} & \multicolumn{1}{a|}{WebSearch} \\ \hline \hline
\multicolumn{2}{|a|}{Disk Capacity} & 16GB & 14GB & 16GB \\ \hline
\cellcolor{Gray} & \multicolumn{1}{a|}{Total Count} & 4.028M & 3.808M & 1.006M \\ \hhline{|>{\arrayrulecolor{Gray}}->{\arrayrulecolor{black}}-|---}
\multirow{-2}{*}{\cellcolor{Gray}Command} & \multicolumn{1}{a|}{Read Ratio} & 12.48\% & 35.50\% & 99.98\% \\ \hline
\cellcolor{Gray} & \multicolumn{1}{a|}{Total Size} & 153.23GB & 30.43GB & 15.24GB \\ \hhline{|>{\arrayrulecolor{Gray}}->{\arrayrulecolor{black}}-|---}
\cellcolor{Gray} & \multicolumn{1}{a|}{Read Ratio} & 5.85\% & 32.73\% & 99.99\% \\ \hhline{|>{\arrayrulecolor{Gray}}->{\arrayrulecolor{black}}-|---}
\cellcolor{Gray} & \multicolumn{1}{a|}{Avg. Read Size} & 17.83KB & 7.36KB & 15.14KB \\ \hhline{|>{\arrayrulecolor{Gray}}->{\arrayrulecolor{black}}-|---}
\multirow{-4}{*}{\cellcolor{Gray}Request} & \multicolumn{1}{a|}{Avg. Write Size} & 40.91KB & 8.33KB & 8.60KB \\ \hline \hline
\end{tabular}
\caption{Characteristics of block I/O traces.}
\label{table:trace}
\end{table}
\end{scriptsize}

\subsection{FTL execution time}\label{sec:time}

\begin{figure*}[!t]
\vspace{-0.8cm}
\begin{center}
\hspace*{-0.4cm} \vspace{-0.6cm}
	\includegraphics[height=0.19\textheight, keepaspectratio=true]{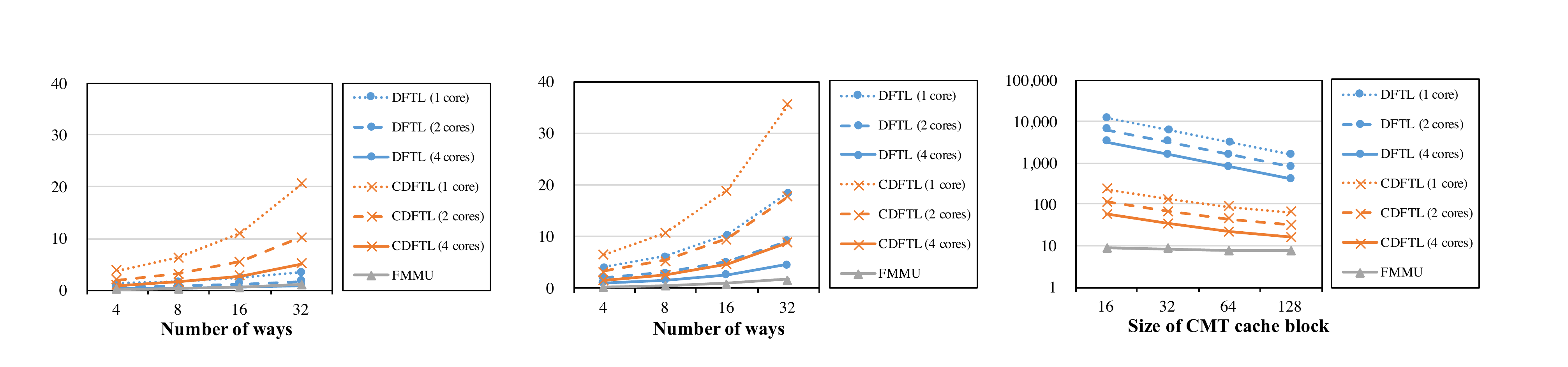}
\begin{flushleft}
\hspace{2cm} (a) map cache hit \hspace{3.1cm} (b) map cache miss \hspace{3cm} (c) map cache flush
 \vspace{-0.1cm}	\caption{FTL execution times of different map cache schemes.} \label{fig:overhead}
\end{flushleft}
 \vspace{-0.2cm}
 \end{center}
\end{figure*}

Figure~\ref{fig:overhead} shows the FTL execution times for different map cache schemes. The execution times of DFTL and CDFTL are measured by increasing the number of 400MHz Cortex-R4 processors to one, two, and four. The FMMU execution time is measured with the clock frequency at 400MHz. The FTL execution times of each map cache scheme are measured in three cases: map cache hit, map cache miss, and map cache flush.


\begin{figure*}[!t]
\begin{center}
\hspace*{-0.5cm}
	\includegraphics[width=1.03\textwidth, keepaspectratio=true]{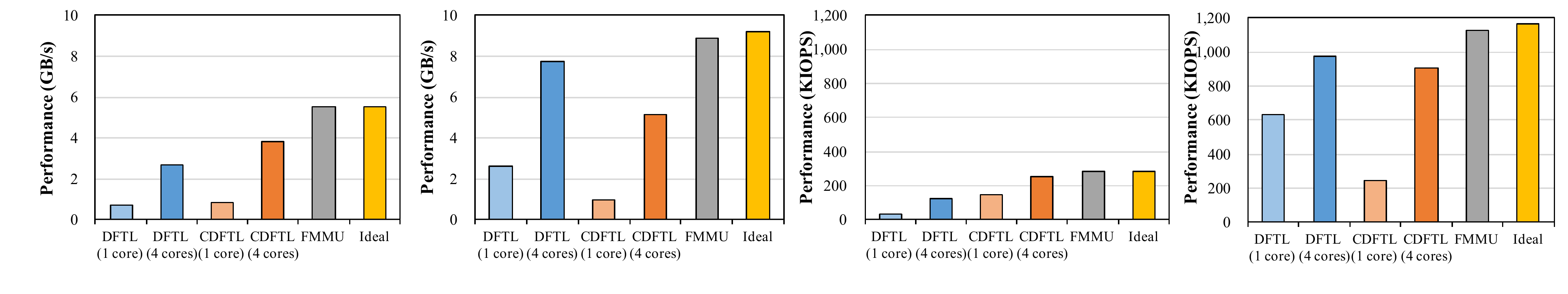}
\begin{flushleft}  \vspace{-0.6cm}
\hspace{0.6cm} (a) 64KB sequential write \hspace{0.7cm} (b) 64KB sequential read \hspace{0.9cm} (c) 4KB random write \hspace{1.2cm} (d) 4KB random read
\end{flushleft}
\vspace{-0.3cm}
	\caption{Performance of different map cache schemes under synthetic workloads.} \label{fig:synthetic}
\vspace{-0.2cm}
\end{center}
\end{figure*}

\begin{figure*}[!t]
\begin{center}
\hspace*{-0.2cm} 
	\includegraphics[width=1\textwidth, keepaspectratio=true]{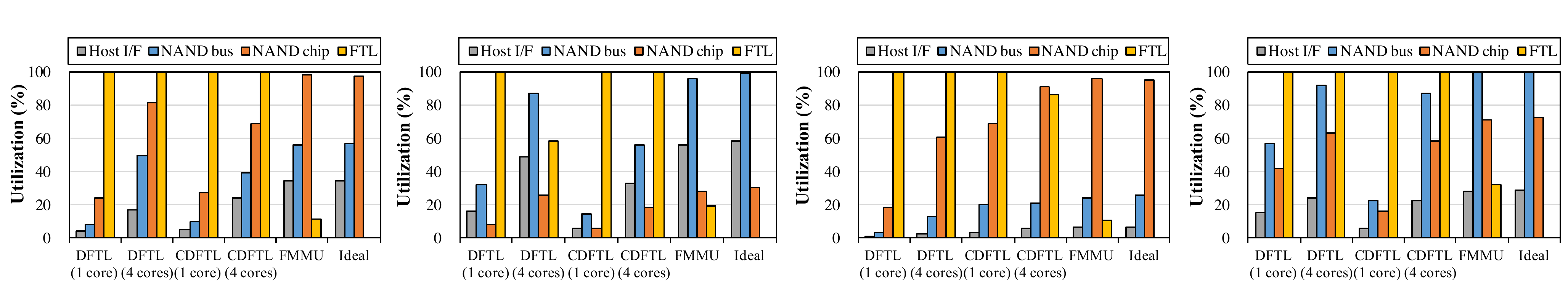}
\begin{flushleft}  \vspace{-0.3cm}
\hspace{0.6cm} (a) 64KB sequential write \hspace{0.7cm} (b) 64KB sequential read \hspace{0.9cm} (c) 4KB random write \hspace{1.2cm} (d) 4KB random read
\end{flushleft}
\vspace{-0.3cm}
	\caption{Utilization of different map cache schemes under synthetic workloads.} \label{fig:synthetic_util}
\end{center}
\end{figure*}

First, the FTL execution time in the map cache hit can be seen in Figure~\ref{fig:overhead}(a). DFTL assumes that cache hit occurs in CMT, and CDFTL and FMMU assume cache miss in CMT, but hit occurs in CTP. In case of 4-way set associative cache, $T_{FTL\_map}$ of CDFTL decreases from 4$\mu$s to 1$\mu$s and $T_{FTL\_map}$ of DFTL decreases from 1.5$\mu$s to 0.4$\mu$s by changing from 1-core to 4-core, respectively. The reason why CDFTL is 2.5 times slower than DFTL is that CMT needs to find the victim cache block by the second chance algorithm and additionally to determine cache hit/miss in CTP. This process occurs in FMMU like CDFTL. However, FMMU processes a request within 0.16$\mu$s since it is implemented as hardware. This result is 5 times and 2.5 times faster than CDFTL and DFTL using 4-core, respectively. As the number of map cache ways increases, the FTL execution time also increases proportionally because the number of tags to be compared increases.


As described in Figure~\ref{fig:overhead}(b), map cache miss and map cache hit show similar results. DFTL is about three times slower than cache hit case because it needs to find victim cache block by second chance algorithm in CMT. In the CTP of CDFTL and FMMU, a victim cache block is selected and a flash read request is issued to the FC to load the translation page from the flash memory. In the DFTL, $T_{FTL\_cmd}$ is measured about in 0.2$\mu$s at 400MHz processor, and this time is about 13\% of $T_{FTL\_map}$. Therefore, in order to reduce the FTL execution time, the execution time of map cache unit which dominates the FTL execution time should be optimized.

 
Finally, Figure~\ref{fig:overhead}(c) shows the result of measuring the execution time of map cache flush. To reduce the number of flash program operations during flush operation, CMT searches for dirty cache blocks belonging to the same translation page and performs batch update. Since DFTL and CDFTL do not provide a data structure to find dirty blocks belonging to the same translation page, all dirty blocks of the CMT must be tested. The flush execution time of DFTL is orders of magnitude slower than that of CDFTL because DFTL uses all 1,088KB of RAM while CDFTL uses 64KB of RAM as CMT. The flush times of DFTL and CDFTL are inversely proportional to the cache block size because if the cache block size of the CMT increases, the number of cache blocks constituting the CMT decreases. FMMU can execute flush within 10$\mu$s because it connects DTL entry and dirty cache blocks belonging to the same translation page through the next link which is included in each cache block.
 

\subsection{Experimental Results}

\begin{figure*}[!t]
\begin{center}
\hspace*{-0.9cm} 
	\includegraphics[width=1\textwidth, keepaspectratio=true]{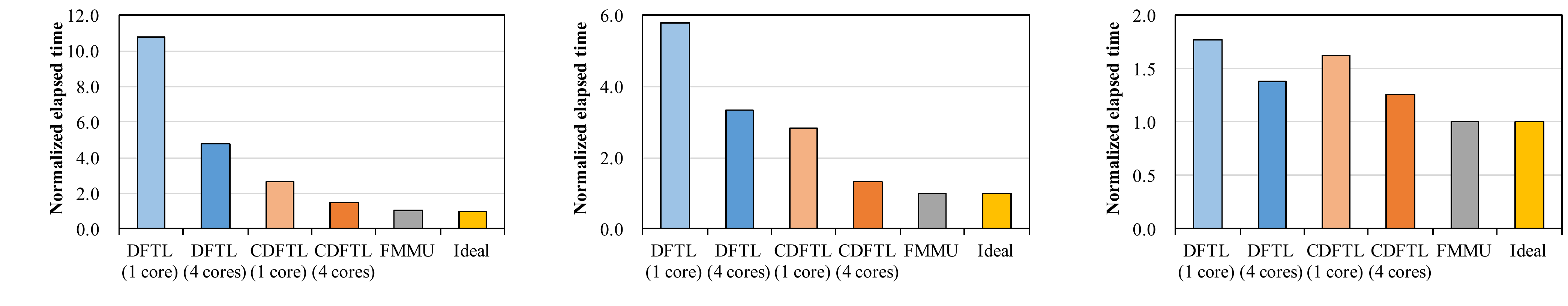}
\begin{flushleft}  \vspace{-0.4cm}
\hspace{2.1cm} (a) MSR\_proj \hspace{3.8cm} (b) MSR\_hm \hspace{3.9cm} (c) WebSearch 
\end{flushleft}
\vspace{-0.4cm}
	\caption{Normalized elapsed time of different map cache schemes under block I/O traces.} \label{fig:trace}
\end{center}
\end{figure*}

\noindent\textbf{Synthetic workload.} As shown in Figure~\ref{fig:synthetic} and Figure~\ref{fig:synthetic_util}, the performance and utilization of SSDs are evaluated under synthetic workload, reflecting the FTL execution time of different map cache schemes measured in Section~\ref{sec:time}. The LBAs of random write and read workload are randomly selected from the entire SSD area. For each test, sequential write is preceded for the whole area of SSD, and random write test is performed until garbage collection is triggered. The ideal case shows the performance when the FTL execution time is set to zero.


First, Figure~\ref{fig:synthetic}(a) shows the performance of 64KB sequential write. The ideal performance is about 5.5GB/s, which is limited by NAND chip performance. In DFTL and CDFTL, the sequential write performance increases as the number of cores changes from one to four since FTL execution time acts as a bottleneck. On the other hand, FMMU shows the performance close to the ideal performance because FTL execution time is shorten by hardware-automated FMMU.


Second, Figure~\ref{fig:synthetic}(b) describes the performance of 64KB sequential read. The ideal performance is about 9.2GB/s, which is limited by NAND bus bandwidth. In DFTL and CDFTL, the NAND bus is under-utilized because the FTL execution time acts as a bottleneck in sequential read performance. DFTL has better performance than CDFTL with 1-core, but it has lower performance than CDFTL with 4-core. Because DTL experiences reading the same translation page multiple times since it does not have a CTP that exploits the spatial locality. As a result, NAND flash memory can not be utilized to the maximum. FMMU achieves the performance close to ideal performance due to CTP and hardware-automation.


Third, 4KB random write performance is illustrated in Figure~\ref{fig:synthetic}(c). Ideal 4KB performance shows about 283KIOPS, and like sequential write, the performance is limited by NAND chips. FTL execution time acts a performance bottleneck in DFTL and CDFTL. DFTL exhibits lower performance than CDFTL because it takes longer flush execution time. CDFTL reaches about 88\% of ideal performance. FMMU uses hardware automation and next link data structure to achieve performance close to ideal performance. FMMU further improves the performance of CDFTL and provides the performance close to ideal performance using hardware automation and next link information.


Finally, Figure~\ref{fig:synthetic}(d) depicts the 4KB random read performance. Ideal 4KB random read performance is about 1,165KIOPS, and it is limited by NAND bus bandwidth. DFTL and CDFTL approach the ideal performance by reducing FTL execution time with 4-core. FMMU allows to achieve the ideal performance, while the FTL utilization is only about 17\%.


\noindent\textbf{Block I/O trace.} As can be seen in Figure~\ref{fig:trace}, the elapsed time of SSD is measured for MSR\_proj, MSR\_hm, and WebSearch trace, and each elapsed time is normalized to the elapsed time of ideal.


DFTL with 1-core under MSR\_proj trace runs 10.63 times slower than ideal as shown in Figure~\ref{fig:trace}(a). Although DFTL can evict multiple dirty cache blocks with a single map flush operation, the DFTL map flush execution time takes orders of magnitude to CDFTL and DFTL. CDFTL with 4-core is still 1.47 times slower than ideal case, but FMMU can approach the ideal performance.


As shown in Figure~\ref{fig:trace}(b), DFTL and CDFTL improve the performance by changing from 1-core to 4-core, but they are 3.35 and 1.32 times slower than ideal, respectively.


Finally, Figure~\ref{fig:trace}(c) presents the performance under WebSearch trace. DFTL exhibits slightly lower performance than CDFTL and FMMU because WebSearch trace includes spatial locality. DFTL generate about 8\% flash read operations more than CDFTL and FMMU.


\subsection{Scalability}

\begin{figure}[!t]
\begin{center}
	\vspace{-0.3cm}
	\includegraphics[width=1\columnwidth, keepaspectratio=true]{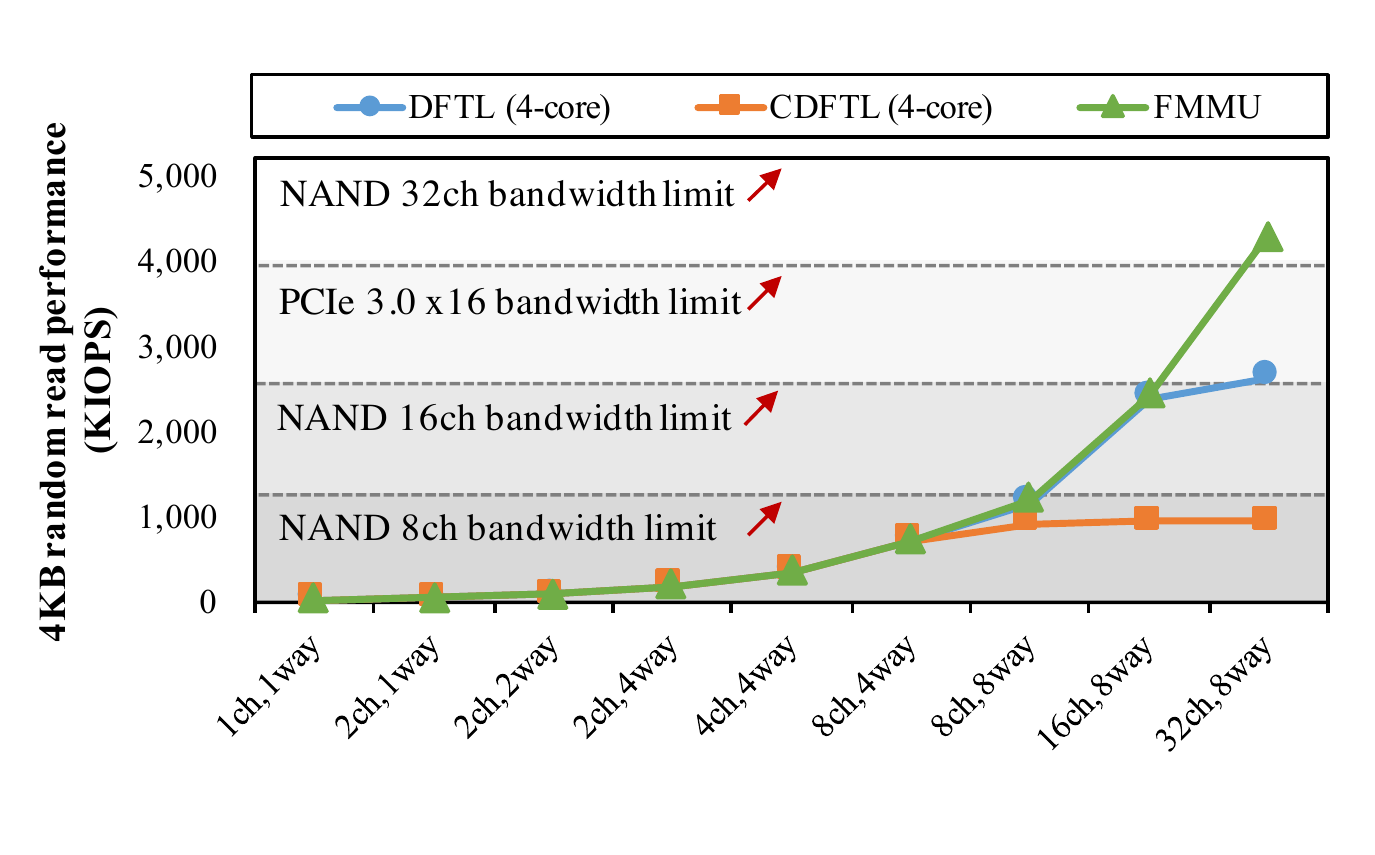}
	\vspace{-1.2cm}
	\caption{Scalability analysis of FMMU.} \label{fig:scalable}
	\vspace{-0.5cm}
\end{center}
\end{figure}

In order to check whether FMMU enables the scalability of the SSD performance, SSD performance is evaluated by varying host interface and NAND configuration with 4KB random read in map cache case. The host interface is used PCIe 3.0 x32, and the NAND configuration is changed from 1-channel, 1-way to 32-channel, 8-way. As can be seen from Figure~\ref{fig:scalable}, SSD with FMMU exhibits about 4.3 MIOPS performance under PCIe 3.0 x32 host interface and 32-channel, 8-way NAND configuration. This performance is limited by the 32-channel NAND bus bandwidth. Therefore, FMMU prevents FTL execution time from being a bottleneck of SSD performance under PCIe 3.0 x32 host interface and 32-channel, 8-way NAND configuration environment.


\section{Related Work} \label{sec:related}

Map cache idea of FMMU is based on previous studies of demand-based page-level mapping FTL. DFTL~\cite{dftl} adopts page-level mapping FTL to reduce garbage collection overhead against block-level and hybrid-level mapping FTL, while uses a small amount of RAM by storing the entire mapping table in the flash memory and by caching only the recently used mapping entries. The DFTL captures temporal locality using the LRU algorithm, but does not exploit spatial locality, so this results in accessing the same translation page multiple times in the sequential workload. CDFTL~\cite{cdftl} adopts the map cache of DFTL as the first-level cache, and uses the second-level cache with the same cache block size as the translation page size to exploit spatial locality. DFTL and CDFTL reduce the number of flash program operations by finding all dirty cache blocks belonging to the same translation page and performing batch update. However, as the number of cache blocks increases, the overhead for finding dirty cache blocks belonging to the same translation page also increases. TPFTL~\cite{tpftl} addresses this problem by adopting two-level LRU lists that manage the translation pages of as an LRU list and introduces two-level LRU lists that manage the translation pages as an LRU list, and cache blocks belonging to the same translation page as an LRU list. These LRU lists are updated for every \textit{Lookup} and \textit{Update} request and must be implemented in software because they are too complicated to implement in hardware. Since the overhead of update LRU lists increases the execution time of the FTL, the execution time of the FTL acts as a performance bottleneck when the performance requirements of the NAND flash-based storage are increased.


FMMU is a hardware-implemented map cache unit that exploits both temporal locality and spatial locality of workload by using first-level and second-level map cache. FMMU also reduces the number of flash programs by performing batch update. In order to minimize the overhead of finding dirty cache blocks belonging to the same translation page, dirty blocks belonging to the same translation page are connected to each other using next link. LRU information is not maintained to the dirty cache blocks because the blocks belonging to the same translation page are evicted together, but the second chance algorithm is applied to the dirty translation pages.


Non-blocking scheme of FMMU is inspired by non-blocking CPU cache algorithms. The Miss Status Holding Register (MSHR)~\cite{mshr}  is a register that stores request information that caused a cache miss. If the CPU cache misses an outstanding cache miss, it sends a request to the memory system and stores the request information that caused the cache miss in a register called MSHR. When a cache miss occurs, the CPU cache sends a request to the memory system and stores the request information that caused the cache miss in the MSHR, so that the CPU cache can continue processing the next request without blocking. When a response arrives from the memory system, the CPU cache finds the corresponding MSHR and sends the data to the destination register of the CPU according to the information recorded in the MSHR. Inverted MSHRs~\cite{inverted} limits the number of MSHR registers by holding the MSHR registers with the same number of destination registers as the CPU. However, this method also has an overhead to find out which MSHR corresponds to the response from the memory system. In-cache MSHRs~\cite{incache} stores outstanding miss information in a cache block allocated to handle cache misses, rather than using a separate register to store outstanding miss information. This method is not only space efficient, but also eliminates the overhead of finding a register when a response is received from a memory system. This method requires a single bit to indicate whether the cache block contains outstanding miss information or contains data. However, since the number of outstanding misses is less than three in a typical CPU cache environment~\cite{cpu_mshr}, requiring one bit per cache block is no longer space efficient.


FMMUs adopts in-cache MSHRs to meet high performance and low RAM usage requirements in environments where many outstanding cache misses occur. The inverted MSHRs approach is not suitable for SSDs because the host system can generate up to 64K outstanding requests and the host sends a number of outstanding requests to maximize the performance of the SSDs. To fully exploit parallelism of the flash memory chips, the map cache unit must be able to handle as many outstanding cache misses as the number of flash chips connected to the SSDs. Therefore, FMMU adopts in-cache MSHRs scheme which is space efficient and does not require additional search overhead.


\section{Conclusion} \label{sec:conclusion}

As the performance of NAND flash memory and host interface are improved, we show the following two phenomena through SSD performance modeling: 1) FTL execution time can act as an SSD performance bottleneck 2) Map cache execution time dominates the FTL execution time. Existing SSDs have shortened the FTL execution time by improving the clock frequency of embedded processors that operate FTL or by increasing the number of embedded processors. However, these approaches have limitations because they increase price, power consumption, or synchronization overhead. We propose Flash Map Management Unit (FMMU) that is implemented the map cache unit by hardware rather than software running on embedded processors. FMMU adopts two-level map caching mechanism and non-blocking scheme of CPU cache. Through the next link, a simple data structure that can be implemented in hardware, FMMU quickly finds mapping entries belonging to the same translation page and performs batch update to reduce the number of flash operations. FMMU employs in-cache MSHRs to space-efficiently process outstanding HRM/GC requests in a non-blocking manner. Our performance evaluation with various synthetic workloads and block I/O traces shows that FMMU can prevent FTL execution time from becoming an SSD performance bottleneck.




\bibliographystyle{IEEEtranS}
\bibliography{refs}
\AtEndDocument{\par\leavevmode}

\end{document}